\documentclass[11pt,twoside]{article}

\usepackage{asp2006}
\usepackage{epsf}
\usepackage{psfig}
\usepackage{lscape}

\begin{document}
\title{High resolution observations of a selection of faint 1.4GHz radio counterparts to optical sources in the HDF-N}   %%% Fill in title
\author{H. Thrall, T. W. B. Muxlow, R.J.Beswick, A. M. S. Richards}   %%% Fill in author names
\affil{Jodrell Bank Observatory, The University of Manchester, Macclesfield, Cheshire, SK11~9DL, UK\\}    %%% Fill in author affiliations

\begin{abstract}We present recent results from very deep MERLIN/VLA 1.4GHz observations of 
an 8.5 square arcminute field centred on the the Hubble Deep Field North, in conjunction 
with BVIz catalogues of GOODS/ACS sources detected in this field. In an extension of work 
by Muxlow {\it et al.} (also presented at this conference) on the statistical properties 
of over 8000 sources, the results presented here take advantage of the deep (rms: 3.3 
microJy/bm), sub-arcsecond radio imaging of this field to characterise the properties at 
1.4GHz of a number of individual radio counterparts to GOODS/ACS sources.
 %%% Abstract to run on from here.
\end{abstract}

\section {Introduction}

The Hubble Deep Field (HDF) has been the subject of a number of deep surveys across the 
observing spectrum. Large-area multiband surveys with the Hubble Space Telascope's 
Advanced Camera for Surveys (ACS), such as the Great Observatories Origins Survey (GOODS) 
\citep{giavalisco2004} have made significant contributions to our understanding of galaxy 
evolution. Such studies demonstrate the utility of high resolution multiband imaging, as 
the multiple wavebands give us acces to photometric redshifts and star formation rates, 
while the high resolution enables the morphological classification of galaxies out to 
distant redshifts.

Deep radio observations of the HDF-North at 1.4GHz were made with the VLA in 1996, with 
initial results presented by \citet{richards1998} and \citet{richards2000}. These VLA 
observations were combined with an 18 day MERLIN observation by \citet{muxlow2005}, to 
produce one of the most sensitive high resolution radio images yet made. A total of 92 
radio sources (down to a completnes limit of 40$\mu$Jy beam$^-$$^1$) were detected in this 
image.

In an extension to the work mentioned above, the MERLIN+VLA data has been further 
analysed, in conjunction with the GOODS/ACS BVIz data catalogues. The statistical 
properties of radio sources associated with catalogued optical galaxies in the HDF-North 
are presented in Muxlow {\it et al.} (in prep.), and also in these proceedings. In this 
proceedings we present a brief overview of the results from Thrall {\it et al.} (in prep), 
where we present the properties of individual faint radio sources associated with optical 
galaxies in the HDF-North. Additional analyses of these radio observations, in conjunction 
with GOODS data at X-ray and infrared wavelengths, are also presented in these proceedings 
by Beswick {\it et al.} and Richards {\it et al.}.

\section {Data}

Detailed descriptions of the observation, calibration and combination of the MERLIN+VLA 
data are given in \citet{muxlow2005} and \citet{richards2000}, so only a very brief 
summary is given here. An area centred on the HDF-North was observed at 1.4GHz with the 
VLA in A-array for a total of 50 hours, and with MERLIN for a total of 18.23 days. 81 
postage-stamp images were created, their centres aligned and they were stiched together 
and flattened, resulting in an 8.5' square image of the HDF-North with an rms noise of 
$\sim$3.3$\mu$Jy. The 4 GOODS ACS images (BVIz) were then astrometrically aligned with 
this 
image.

\subsection{Sample Selection}

Taking the GOODS/ACS z-band catalogue of sources (data release v1.0) as the starting 
point, sources were discarded if they lay outside the MERLIN+VLA field. To remove 
confusing sources, pairs or groups of sources separated by less than 1.5 arcsec were 
removed, along with any sources within 3.0 arcsec of the positions of the 92 known bright 
radio sources in the field (Muxlow {\it et al.} 2005). The resulting list of sources was 
matched with the ACS V, I and B-band catalogues, to give a final list of 8403 sources with 
4-band photometry. The statistical properties of these sources are discussed in Muxlow 
{\it et al.}(in prep).

The AIPS task IRING was run on the radio map at the position of each source in the list, 
giving the average and cumulative flux in a series of concentric annuli from 0.25 to 
1.5arcsec in radius, in increments of 0.25arcsec.
 159 of the brightest sources selected by greatest radio flux contained within 0.75arsec radius were chosen for more detailed study. This final sample of 159 faint radio sources is the focus of this proceedings and of Thrall {\it et al.} (in prep).

\section{Source properties}
\subsection{Redshifts}
Spectroscopic redshifts have been published for a small number of HDF sources, but for the vast majority of the sources in the field we have to resort to the less accurate technique of photometric redshift determination. \citet{benitez} has developed a reliable method for determining photometric redshifts using Bayesian inference, using the HDF-N as its testing ground, with excellent results. The Bayesian photometric redshift estimation code (BPZ) described by \citeauthor{benitez} is used here to determine redshifts.

A search of the literature yielded published redshifts for 60 of the 159 sources in this sample. To supplement this list, photometric redshifts were obtained using the BPZ code with certain criteria applied. For a complete discussion of this, see Muxlow {\it et al.} (in prep.). Figure 1 (left) shows the redshift distribution of the 95 sources in this sample for which reliable redshifts have been found.

\begin{figure}
\plottwo{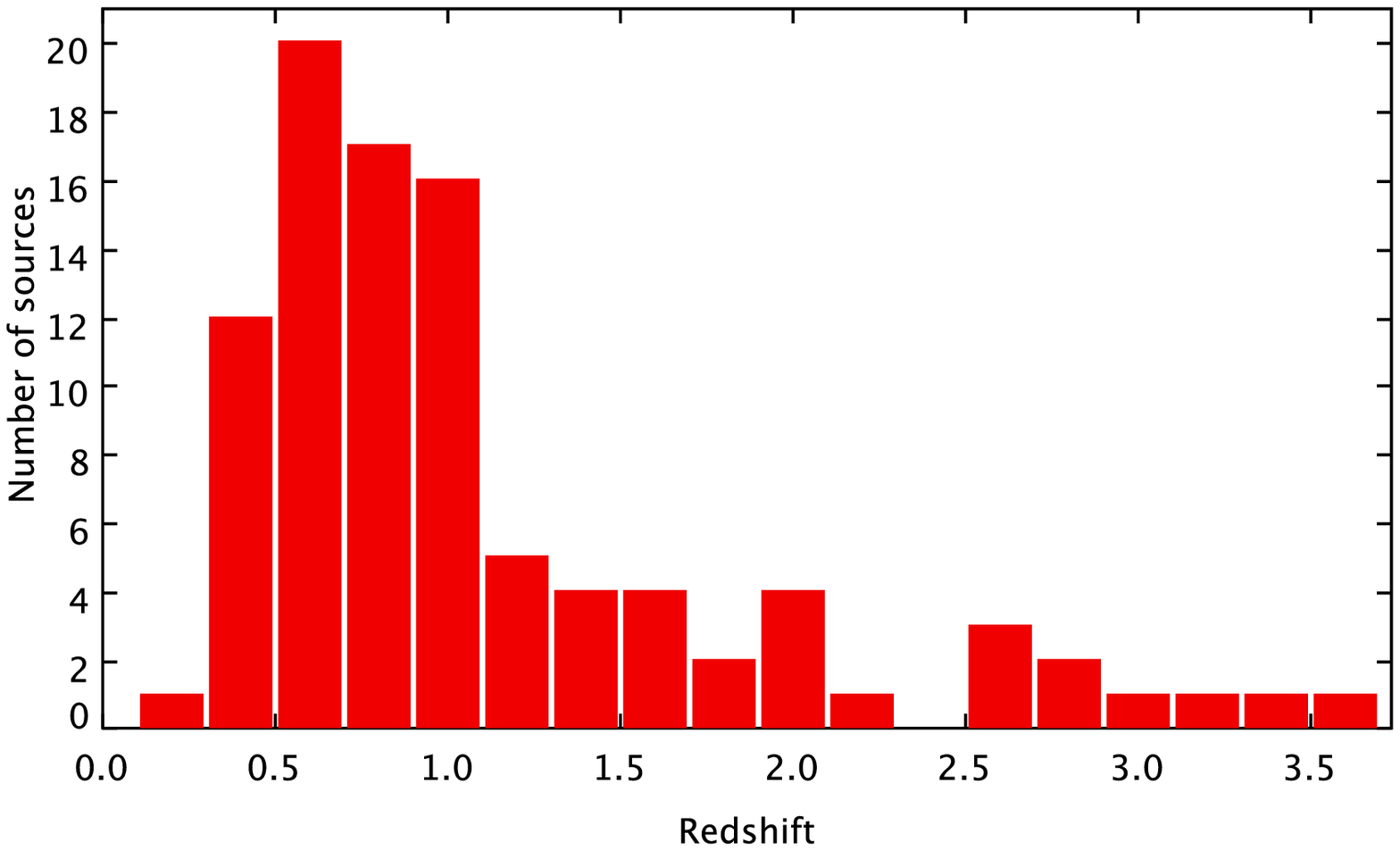}{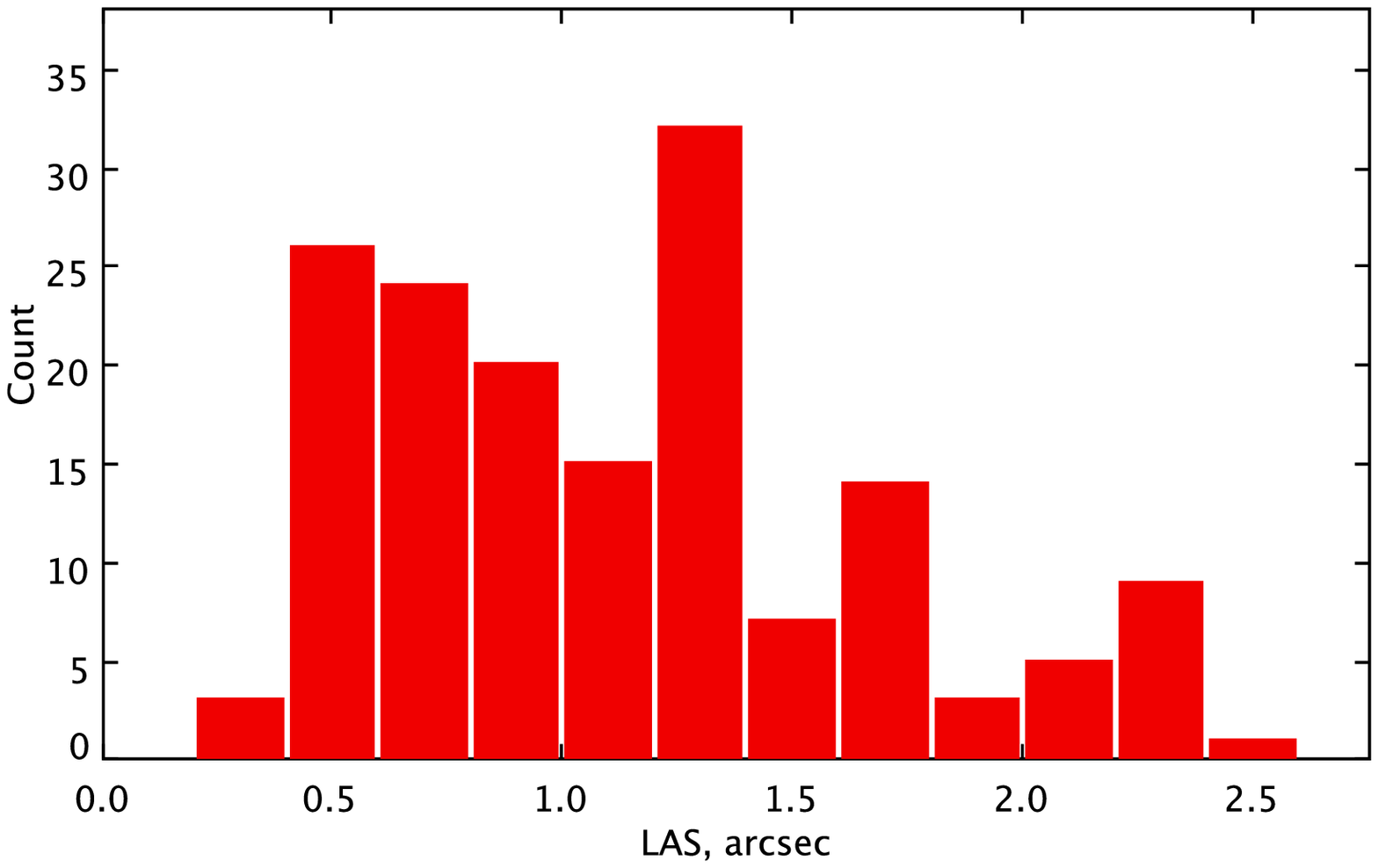}
\caption{(Left) Redshift distribution for the 95 sources out of 159 for which reliable redshifts have been estimated. (Right) Distribution of largest angular size (LAS) for all 159 sources discussed in this paper. These sizes are either calculated from gaussian fits to simple sources, or estimated by visual inspection.}
\end{figure}

\subsection{Sizes}
The sizes of each of the 159 sources in this sample have been estimated. For those with a simple radio structure it was possible to use Gaussian fitting to a slice across the source, while for those with a more complex structure, estimates were obtained by visual inspection. Figure 1 (right) shows the distribution of estimated largest angular size of the 159 sources.

\section{Starbursts and AGN}
Where possible, sources in this sample have been classfied as starbursts or AGN. Candidates for the starburst category have radio emission extended on sub-galactic scales and are also detected in the infrared by Spitzer (ref). AGN candidates have a central radio core along with double-sided emission features. 
In this section four individual sources are presented, as examples of those classified as starbursts and AGN.

\begin{figure}
\plottwo{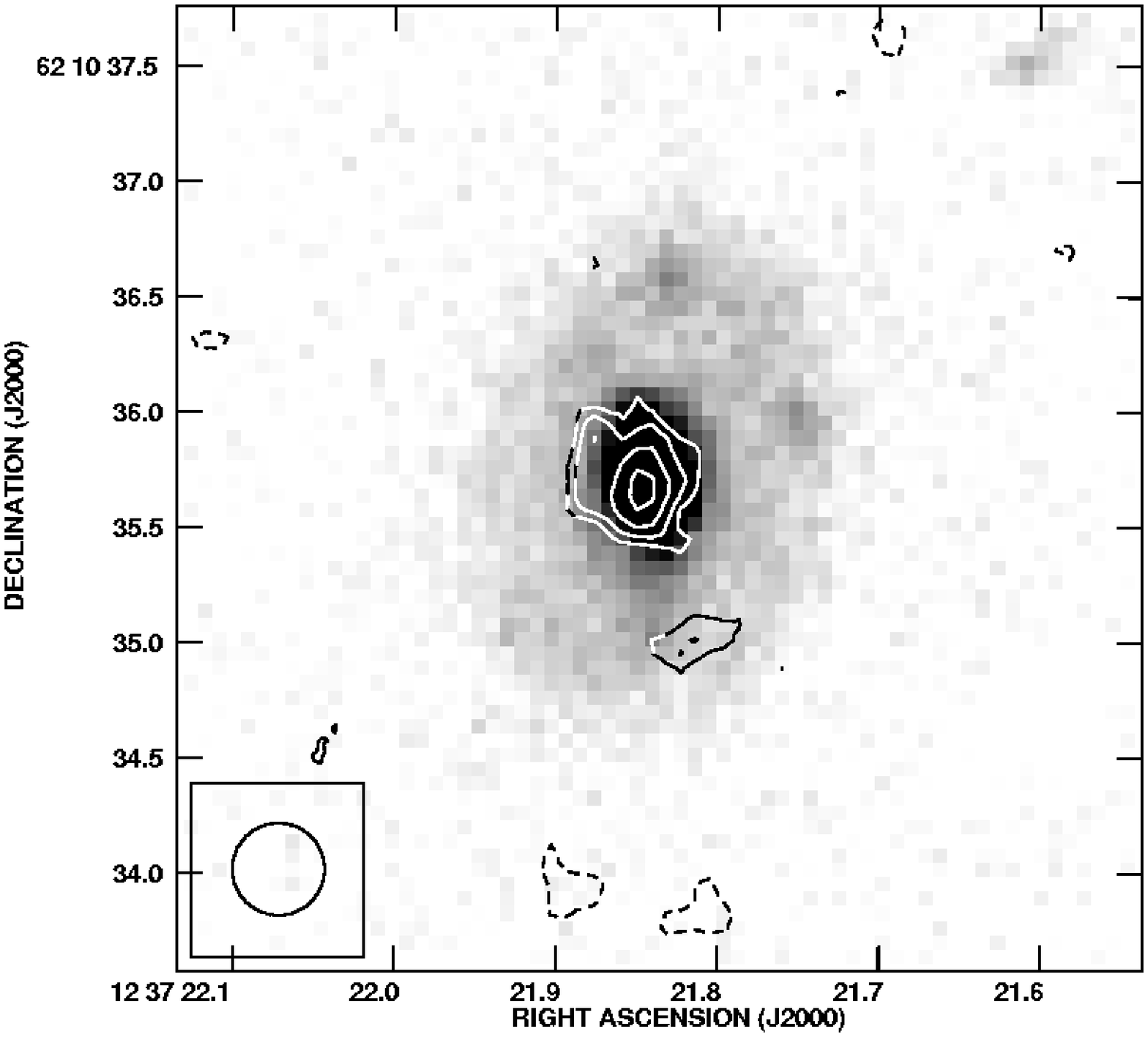}{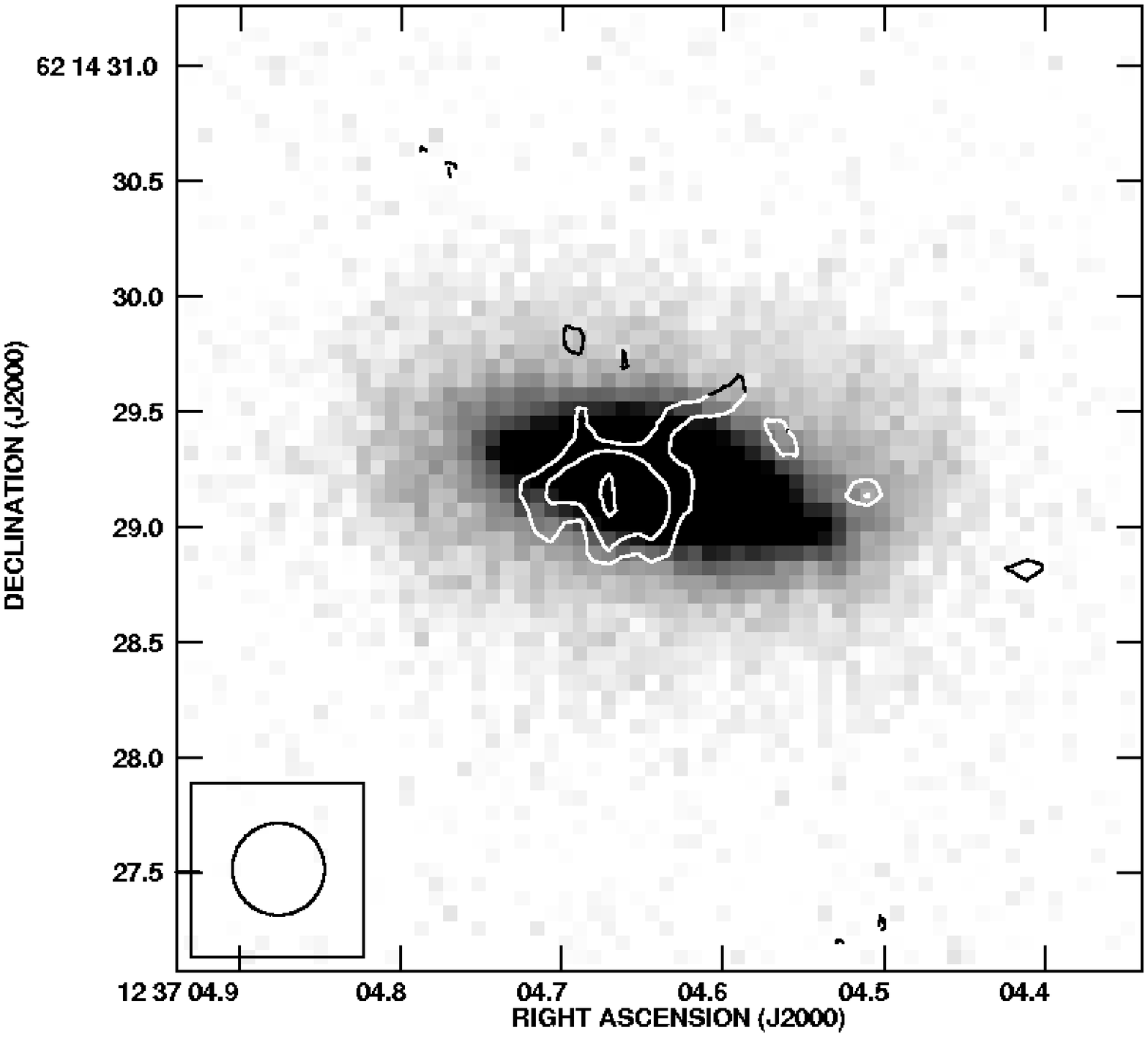}
\caption{(Left) Starburst candidate source J123721.84+621036.0. (Right) Starburst candidate source J123704.64+621429.5. 1.4GHz radio contours overlay 4-colour ACS image.}
\end{figure}

\begin{figure}
\plottwo{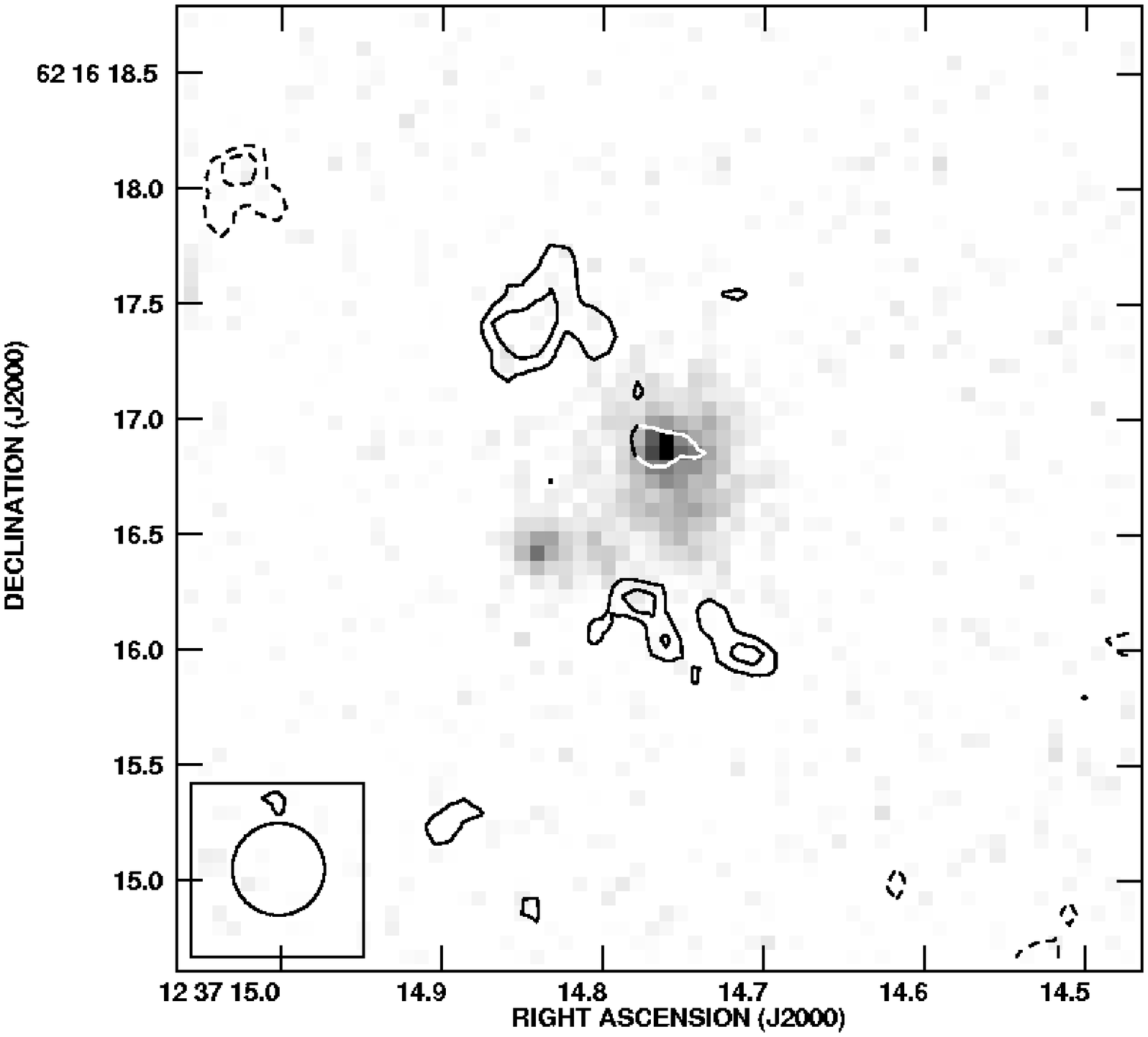}{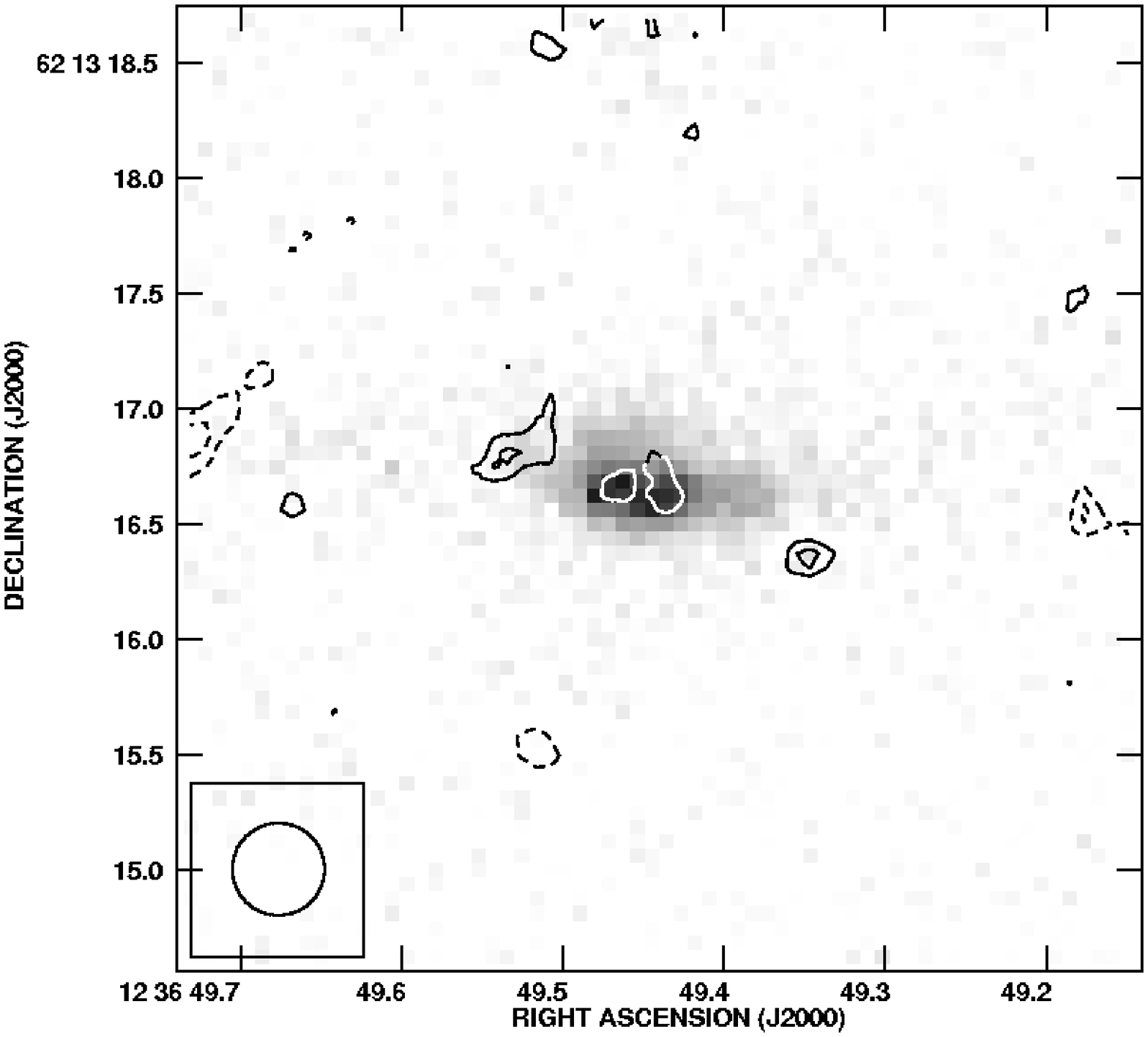}
\caption{(Left) AGN candidate source J123714.77+621617.0. (Right) AGN candidate source J123649.44+621317.0. 1.4GHz radio contours overlay 4-colour ACS image.}
\end{figure}

{\bf J123721.84+621036.0} (see Fig.~2) is a starburst candidate which shows compact radio emission at the centre of the optical galaxy, which has {\it z} = 21.2 mag and redshift 1.0. The angular extent of the central radio emission is 0.5 arcsec, corresponding to a size of $<$5kpc. This source is detected at 24$\micron$ by Spitzer, and in X-rays by Chandra.

{\bf J123704.64+621429.5} (see Fig.~2) is a starburst candidate with compact radio emission lying within the central region of the optical galaxy, which has {\it z} = 20.6 mag. The galaxy is at redshift 0.56 and the radio emission is extended over 1.2 arcsec, which corresponds to a size of $\sim$8kpc. This source is also detected at 24$\micron$ by Spitzer, and in X-rays by Chandra.

{\bf J123714.77+621617.0} (see Fig.~3) has a classic AGN structure. The compact core overlies the nuclear region of the optical galaxy at redshift 1.5 with {\it z} = 23.3 mag. Radio emission on two sides extends over 2.3 arcsec, which at this redshift equates to $\sim$18kpc.

{\bf J123649.44+621317.0} (see Fig.~3) has a radio structure typical of an AGN. The compact core lies at the centre of the optical galaxy, which has {\it z} = 22.8 mag and lies at redshift 1.2. The two-sided radio emission is extended over 1.6 arcsec, which corresponds at this redshift to a size of $\sim$12kpc

%%% MAIN BODY OF TEXT GOES HERE. CONSULT "INSTRUCTIONS FOR AUTHORS USING
%%% LATEX2E MARKUP", SECTIONS 2.3-2.6 FOR HELP WITH EQUATIONS, FIGURES,
%%% AND TABLES.

%\section{}   %%% Top level section head (remove "%" symbol)
%\subsection{}   %%% Second level section head (remove "%" symbol)
%\subsubsection{}   %%% Lowest level section head (remove "%" symbol)
%\section*{}    %%% Unnumbered top level section head (remove "%" symbol)
%\subsection*{}   %%% Unnumbered second level section head (remove "%" symbol)

\acknowledgements %%% Text of acknowledgements runs on after this command.
HT acknowledges a PPARC studentship. RJB acknowledges financial support by the European Commission's I3 Programme ``RADIONET'' under contract No.~505818. Based on observations made with MERLIN, a National Facility operated by the University of Manchester at Jodrell Bank Observatory on behalf of PPARC, and the VLA of the NRAO is a facility of the NSF operated under cooperative agreement by Associated Universities, Inc.

\end{document}